\documentclass [12pt]{article}
\usepackage{times}
\usepackage{natbib}
\usepackage{graphics}
\usepackage{color}
\newcommand{\cmtwo}{cm$^{-2}$}
\newcommand{\cmthree}{cm$^{-3}$}

\newcommand{\kms}{km\,s$^{-1}$}       %km/s
\newcommand{\ecsarc}{erg cm$^{-2}$ s$^{-1}$ arcsec$^{-2}$}

\newcommand{\um}{$\mu$m}                                 %micron
\newcommand{\molh}{H$_{2}$}                              %H_2, H_2O and H II

\newcommand{\water}{H$_{2}$O}

\newcommand{\lsun}{$L_{\odot}$}                          %solar and terr. units

\newcommand{\rsun}{$R_{\odot}$}

\newcommand{\mearth}{$M_{\oplus}$}

\newcommand{\mmoon}{$M_{\rm Moon}$}
\newcommand{\mjup}{$M_{\rm Jupiter}$}
\newcommand{\gapprox}{$\stackrel {>}{_{\sim}}$}   %greater/less approx.
\newcommand{\lapprox}{$\stackrel {<}{_{\sim}}$}
\newcommand{\about}{$\sim$}                       %approx
\newcommand{\powten}[1]{10$^{#1}$}
\newcommand{\ctwo}{[C\,{\sc ii}]\,157\,$\mu$m}    %fine structure lines

\newcommand{\bpic}{$\beta \, {\rm Pic}$}          %sources
\newcommand{\epseri}{$\varepsilon \, {\rm Eri}$}

\newcommand{\asec}{$^{\prime \prime}$}
\newcommand{\adeg}{$^{\circ}$}

\newcommand{\asecdot}[2]{\mbox{#1$\stackrel {\prime \prime}{_{\bf \cdot}}$#2}}

\begin{document}

\begin{center}
{\bf Gas in Dusty Debris Disks}
\end{center}

\begin{center}
Ren\'e Liseau\footnote{Collaborators on this project include 
Pawel Artymowicz, Alexis Brandeker, G\"oran Olofsson (Stockholm 
Observatory, Sweden), Malcolm Fridlund (ESA-ESTEC, The Netherlands) 
\& Taku Takeuchi (University of California, Santa Cruz, USA). 
However, the responsibility for this article rests entirely with the author.}
\end{center}

\begin{center}
Stockholm Observatory, AlbaNova University Center, 
SE-106 91 Stockholm,
Sweden, {\it rene@astro.su.se}
\end{center}


\begin{abstract}
The presence of gas in dusty debris disks around main-sequence stars is reviewed.
We present new observational results for the most prominent representative of 
the class, viz. the southern naked-eye star \bpic toris. The spatial and spectral
distribution of observed atomic lines from the disk around the star is
reproducable by a Keplerian rotation model to a high degree of accuracy. The 
expected velocity dispersion due to radiation pressure in resonance lines
is not observed. Modeling the motion of different atomic species
under the influence of gravity, radiation pressure and gas friction leads to
the conclusion that an underlying decelerating component must be present in the disk.
This braking agent is most likely hydrogen, with inferred average densities 
$n_{\rm H}> 10^6$\,\cmthree. This could support the observational result of 
Thi et al. (2001) which indicated the presence of appreciable amounts of \molh\ 
around the star \bpic.
 
\end{abstract}

\section{Introduction}

One of the major highlights of the IRAS mission was the discovery of
the `Vega-excess' phenomenon exhibited by a number of main-sequence stars
(Aumann et al. 1984), i.e. emission in excess
over the stellar photospheric value at long wavelengths (\gapprox\,10\,\um).  
Already from the very beginning it was recognised that this excess 
is re-emitted stellar radiation
by a surrounding cloud of dust particles. The image of the southern
Vega-excess star \bpic\ in the optical by Smith \& Terrile (1984) 
confirmed the earlier notion that the spatial distribution of the 
dust is highly flattened, i.e. in a disk shaped configuration rather 
than in a spheroidal cloud. Recent reviews of \bpic\ and similar systems - 
the dusty debris disks - include those by Artymowicz (2000), 
Lagrange et al. (2000) and Zuckerman (2001).

\subsection{Dusty Debris Disk Evolution}

Besides Vega ($\alpha$\,Lyr) and \bpic, other prominent members of the class are  
Fomalhaut ($\alpha$\,PsA), \epseri\ and HR\,4796A. The excess emission of these
objects seems to outline an evolutionary trend (Spangler et al. 2001, Fig.\,1),
in the sense that the amount of {\it emitting} particulate matter in a given object declines 
with time. Yet younger objects would occupy the upper left corner of the diagram 
(with considerable spread due to increasing optical depth) and it seems reasonable 
to assume that the dusty debris disks (somehow) are linked to those of the pre-main-sequence 
(PMS) phase. PMS-disks, with ages \lapprox\,10\,Myr, are essentially gaseous 
and are believed to provide the material necessary for the formation of planetary systems 
\cite{calvet2000}. Observed asymmetries and inhomogeneties in the dust emission from
debris disks have been taken to provide evidence for dynamical
signatures of disk-planet 
interaction, (e.g., Ozernoy et al. 2000, Augereau et al. 2001, Wyatt \& Dent 2002, 
Kuchner \& Holman 2003). 

\subsection{Planets in Dusty Debris Disks}

Our knowledge regarding the incidence of {\it known} planets (gas giants)
around debris disk stars is 
very limited. Heavy observational bias is at work here, though. Radial-velocity 
studies are biased toward relatively late spectral types, whereas the most prominent
debris disk stars are of earlier spectral types. In addition, radial-velocity 
studies have focused on fairly old stars, whereas disks around stars much
older than about 400\,Myr would generally not have been picked up in the IRAS survey
\cite{habing1999}.
Three planetary systems, discovered with the radial-velocity technique,
have been claimed by Trilling et al. (2000)
to possess dust disks. The observational evidence presented by Jayawardhana et al. (2002) 
(see also Schneider et al. 2001)
argues strongly against any substantial amounts of dust around 55\,Cnc, whereas
the other two systems, viz. $\rho$\,CrB and HD\,210277, still need confirmation. 
Pantin et al. (2000) reported the detection of a circumstellar disk around $\iota$\,Hor. 
However, this result has not yet been published in a refereed journal.

The {\it circumstantial} evidence for planet(s) around dusty debris stars is enormous.
To give just a few examples, the case for
Vega has been presented by Wilner et al. (2002), for Fomalhaut by Holland et al. (2003) and
for \bpic\ by Wahhaj et al. (2003). For the chromospherically active star
\epseri, Hatzes et al. (2000) presented a rather noisy radial velocity curve (for 
more recent data, see http://exoplanets.org/esp/epseri/epseri.shtml) and the confirmation
of the planet(s) would need a longer baseline in time. 

\subsection{Widespread Gas in Dusty Debris Disks}

If the evolutionary hypothesis is correct, then a similar apparent decrease in the gas content 
may be expected and one could hope to find objects being currently 
in a transitional phase between gas-rich PMS- and gas-poor debris disks. This could
open up the possibility to directly witness the end phases of planet formation and 
the early stages of planetary system evolution. Appreciable amounts of gas could still 
be present in these systems.

Alternatively, in older systems, where the dust debris is most likely of more recent origin,
the study of the dust production mechanism(s) could benefit from complementing observations
of the accompanying gaseous components, 
as for example in the case of solar system comets \cite{cremonese2002}. 

In either of these scenarios, the detection of sharp spectral features originating from the gas 
would permit the direct observational study of the disk kinematics and provide input to 
theoretical models of the debris disk dynamics.

\section{Searches for Molecular Gas}

Recalling that the discoveries were made at IRAS wavelengths, it was clear
already from the very beginning that the emitting dust grains had to be at relatively low 
temperatures (a few tens to some hundred Kelvin). The emission being optically thin, these
observations led directly to estimates of the total amount of dust (particle sizes 
\lapprox\,$\lambda$) in these systems \cite{hildebrand83}. 
Disregarding for a moment the many details which in reality
enter these estimations such as, e.g., the correct description of the dust absorption 
coefficient(s) \cite{beckwith2000} - the mass is inversely proportional to this quantity - 
inferred lower limits to the dust mass are some fraction to some tens of a lunar mass 
(\mmoon). For comparison, the Minimum Mass Solar Nebula amounts to roughly $5 \times 10^4$\,\mmoon, 
from which it is clear that, for any reasonable assumption about the local gas-to-dust mass ratio, 
any solar system type of planet formation must have been completed by the Vega excess stage.

If the gas-to-dust mass ratio in these systems were comparable to that of the interstellar
medium (ISM, \about\,160), detectable amounts of (primarily molecular) gas could be expected. In
fact, carbon monoxide (CO) has frequently been observed through disk absorption against the 
ultraviolet stellar continuum of \bpic\ (Roberge et al. 2000 and references therein). In an
optical absorption study, Hobbs et al. (1985) put an upper limit on the amount of CH$^+$
in the \bpic\ disk.

Emission line studies, mostly in the sub-/millimeter regime, have so far been essentially unsuccessful 
(Liseau 1999 and references therein).
Molecules searched for in emission include CO, CS, SiO and HCO$^+$ from the gound and
\water\ and O$_2$ (the SWAS Team) from space. In addition, the detection of rotational
\molh\ emission in the mid-infrared (ISO-SWS) from \bpic\ has been reported \cite{thi2001},
with an estimate of the total mass of the order of 50\,\mearth. 
However, Lecavelier des Etangs et al. (2001) were unable to confirm these amounts 
of \molh\ through absorption 
measurements in the far ultraviolet (FUSE) and the matter seems at present inconclusive
(but see below, Sect.\,3.1.2).

Various explanations have been offered to explain the observed low levels of molecular 
concentration in dusty debris disks, particularly what regards CO, being ubiquitous
in the ISM and in solar system comets. These include `abnormally' low gas-to-dust mass 
ratios, `abnormally' low {\it gas} phase CO abundances (relative to \molh), 
photodissociation of CO and, simply, non-existence (early dissipation 
through stellar wind and/or consumption during planet building). None of these explanations is 
flawless and the resolution of this issue will have to await future observational and 
theoretical improvements.

\section{Searches for Atomic Gas}

Observational studies of atomic gas, either ionised or neutral, in dusty debris disks
have traditionally focused on line {\it absorption} in a very limited number of objects.
Although very sensitive to column densities, these studies are generally incapable of
pin-pointing the regions of line absorption along the line of sight.
In the following, a number of observational studies are discussed, which were aiming at
the detection of {\it emission} from atomic gas.

\subsection{Atomic Gas Emission from the \bpic\ Disk}

The by-far best studied object among the dusty debris disks is the \bpic\ system. The attempt to
directly measure, through 21\,cm line emission, the amount of atomic hydrogen gas was unsuccessful
\cite{freudling1995}. The presence of widespread ionised hydrogen in the disk is not expected
and Balmer line emission has not been detected \cite{brandeker2002}. 
Observations with the ISO-LWS may have detected feeble \ctwo\ emission toward \bpic\ 
\cite{kamp2003}. Few details about the circumstellar disk could be learned from these
spatially and spectrally under-resolved observations, though. 

The resolution issue was greatly improved upon by using a large optical telescope,
equipped with an Echelle spectrograph and a long slit, leading to the discovery of widespread 
sodium gas in the disk \cite{olofsson2001}. These observations provided finally conclusive
evidence that the star \bpic\ is indeed surrounded by a disk in Keplerian rotation, seen
nearly edge-on (Fig.\,2).

Follow-up observations \cite{brandeker2003}
covered the optical spectrum (0.3 to 1.1\,\um) at greater resolving power
and, in particular, covered a larger area with the spectrograph slit positioned both along
and perpendicularly to the disk-midplane (Figs.\,2 and 3). 
Projected slit widths were 
6\,AU and slit lengths 154\,AU, covering the disk out to 300\,AU in both the northeast (NE) and
southwest (SW) parts of the disk. Perpendicular slits were positioned at approximately 
60\,AU and 120\,AU on both sides from the star, extending to 80\,AU on either side of the
midplane (Fig.\,3). 

Nearly 80 emission lines from the disk were detected, all from non-volatile elements
including Ca\,II, Cr\,I+II, Fe\,I, Na\,I, Ni\,I+II, Ti\,I+II. In addition, a dozen lines 
lack so far a clear identification. Neutral iron, with more than 50 strong
lines, dominates the disk spectrum. Measurements of the line emission along the perpendicular
slits revealed that the gas disk is not perfectly plane (Fig.\,4). Inside
$\pm 3$\asec, the disk appears tilted at an angle of about 5\adeg, very much the same as
inferred for the dust disk by Heap et al. (2000). Further out, the gas disk is flaring at
an opening angle of about 10\adeg\ ($H/r \sim 0.2$).

The two disk halves display significant asymmetries and small scale structure in their
line emission, as illustrated by NaI\,D$_2$ in Fig.\,5. The NE disk extends much
farther out, in fact to the limit of our observations (cf. Fig.\,2). In
contrast, further in, the SW side is brighter. These features can also be
discerned in the dust disk, albeit at lower contrast \cite{heap2000}.

\subsubsection{The Kinematics of the Atomic Gas Disk}

As a complete surprise came the observed small velocity dispersion of the line emission
from the atomic disk \cite{olofsson2001}. In the absence of a major component of volatile gases,
an excess of some hundred of \kms\ over the local Keplerian value of the radial velocity
would be expected due to radiation pressure on the neutral sodium atoms. 
As different species will be differently
sensitive to radiation pressure, their spectral lines should exhibit different widths. 
Below, the radiation pressure coefficient is calculated for various atoms/ions
observed in the \bpic\ disk.

For a single transition $i$, the radiation pressure coefficient $\beta_i$, expressing the 
relative importance of radiation pressure and gravity, can be written as

\begin{equation}
\beta_i = \frac{F_{\rm rad}}{F_{\rm grav}} 
      = \frac{\pi\,r_{\rm e}\,f_{\rm l}\,\lambda^2\,F_{\lambda}(R_{\star})/\,c}
             {10^{\log{g_{\star}}}\,{\rm A_{atom}}}\,\,\,,
\end{equation}

where the flux density $F_{\lambda}$ refers to the stellar photosphere at the wavelength 
$\lambda$ of the line and the gravity is given by the surface $\log{g_{\star}}$ and the 
atomic mass number ${\rm A_{atom}}$ of the ion, expressed in amu, and where

\begin{equation}
\pi\,r_{\rm e}\,f_{\rm l} = \frac{1}{8\,\pi} \frac{g_{\rm u}}{g_{\rm l}} 
                            \lambda^2 A_{\rm ul}\,\,\,.
\end{equation}

\begin{table}
  \begin{center}
    \caption{Radiation pressure coefficients$^{\,\,a}$ $\beta$ for the \bpic\ gas disk;
    ions in alphabetical order}
    \begin{tabular}[h]{lcccc}
      \hline
      Ion    & Vacuum             &  $A_{\rm ul}$ & $\beta_{\rm min}^{\,\,b}$ & $\beta_{\rm max}^{\,\,b}$ \\
             & Wavelength (\AA)   &  (s$^{-1}$)   &                   &                   \\
      \hline
      \\
      Ca\,I  & 4227.918           & 2 .18e+8      &  35               &  270  \\
      Sum (9)$^c$&                &               &  35               &  270  \\
      Ca\,II & 3934.777           &  1.47e+8      &  2.0              &  95   \\
             & 3969.591           &  1.4 e+8      &  1.0              &  34   \\
      Sum (4)&                    &               &  3.1              & 130   \\
      Cr\,I  & 3579.705
          &  1.48e+8      &  3.8              &  16   \\
             & 3594.507           &  1.50e+8      &  3.1              &  13   \\
      Sum (18)&                   &               & 14                &  69   \\             
      Cr\,II & 8002.280           &  1.0 e-1      &  1.e-7            & 1.e-7 \\
      Sum (5) &                   &               &  0                & 0     \\       
      Fe\,I  & 2484.021           &  4.9 e+8      &  0.16             &  11   \\
             & 3861.006           &  9.7 e+6      &  1.3              &  20   \\
      Sum (40)&                   &               &  3.7              &  32   \\                   
      Fe\,II & 2382.765           &  3.8 e+8      &  0.03             &   7   \\
             & 2600.173           &  2.2 e+8      &  0.4              &   5   \\
      Sum (14)&                   &               &  0.6              &  16   \\        
      H\,I   & 1215.6682          & 6.265e+8      &  3.e-5            & 2.e-2 \\
             & 1215.6736          & 6.265e+8      &  2.e-5            & 1.e-2 \\
     Sum (10)&                    &               &  0                & 0     \\                
       Na\,I & 5891.583           &  6.22e+7      & 160               & 170   \\
             & 5897.558           &  6.18e+7      &  80               &  83   \\
      Sum (32)&                   &               & 240               & 254   \\            
       Ni\,I & 2290.690           &  2.1 e+8      & 0.06              &   2.3 \\
             & 2320.747           &  6.9 e+8      & 0.06              &  12   \\
      Sum (36)&                   &               & 0.5               &  20   \\            
      Ni\,II & 1751.910           &  4.8 e+7      & 0.06              &   0.2 \\
      Sum (25)&                   &               & 0                 &   0.2 \\       
             &                    &               &                   &       \\  
      Ti\,I  & 3636.499           &  8.04e+7      & 2.0               &  10   \\
             & 3982.887           &  3.76e+7      & 1.6               &  10   \\
      Sum (40)&                   &               & 14                &  77   \\
             &                    &               &                   &       \\ 
      Ti\,II & 3384.740           &  1.09e+8      & 3.0               &  12   \\
      Sum (7) &                   &               & 3.9               &  19   \\                                      
      \hline \hline \\
      \end{tabular}
    \label{rad_press}
  \end{center}
$^a$ NIST atomic data; model is $T_{\rm eff} = 8000$\,K,
     $\log{g}=4.5$, $\log{Z/Z_{\odot}}=0.0$ \cite{hauschildt1999}.  \\
$^b$ $\beta = F_{\rm rad}/F_{\rm grav}$ and {\it min} and {\it max} values refer to the central
     line absorption and neighbouring continuum, respectively.  \\
$^c$ Number of ground state transitions for total $\beta$.   
\end{table}

$r_{\rm e}$ is the classical electron radius, $f_l$ the absorption oscillator strength and the 
other symbols have their usual meaning. 
For a particular atom/ion, the radiation pressure coefficient is $\beta = \sum_i \beta_i$.
Both $F_{\rm grav}$ and $F_{\rm rad}$ are
inverse square laws so that their ratio is independent of the distance to the star.

In Table\,\ref{rad_press}, $\beta$-values for a number of atoms/ions are presented. These 
are all for transitions from the ground, as the disk gas is too cool to maintain significant 
excitation in the higher states. For the photospheric spectrum the Next Generation Atmosphere
Models were used \cite{hauschildt1999}, which is shown in Fig.\,6. 
Two values of $\beta$ are given per resonance line.
$\beta_{\rm min}$ corresponds to the scattering of a stellar photon by an atom which is
essentially at rest (with respect to the radial velocity component), i.e. the
source is the core of the photospheric absorption line. As the atom picks up speed, a greater number 
of source photons becomes available from the  wing of the photospheric line and 
$\beta_{\rm max}$ refers thus to the neighbouring stellar continuum. In addition, 
as we did not spin up the model atmosphere to the rotation velocity of the star, these
two $\beta$ values respresent truly strict limits to the realistic case. 
Obviously, the differences can be
quite dramatic, as for e.g. the H and K lines of Ca\,II (Table\,\ref{rad_press}).
In contrast, the photospheric Na\,D absorption is quite shallow and, hence, 
$\beta_{\rm min} \sim \beta_{\rm max}$ so that the sodium atoms should be sailing 
at their terminal speed at velocities much higher than what is actually
observed (Olofsson et al. 2001, Brandeker et al. 2003).

\subsubsection{The Dynamics of the Disk Gas}

Also quite obvious from Table\,\ref{rad_press} is the fact that lines originating from different
species and/or different ionisation stages should behave quite differently. 
For instance, the lines of singly ionised 
chromium should reflect only Keplerian rotation, whereas neutral chromium should 
possess a significant radial component. This is not observed \cite{brandeker2003}. 
In fact, regardless of their $\beta$ value, none of the species shows significant
radial velocity excess over the Keplerian value.

A possible explanation could be the existence of considerable amounts of, yet to be 
identified, `quiescent' material, braking the radiatively accelerated gas
to the observed relative velocities. 
An obvious candidate would be hydrogen, being hard to get at observationally.
The star produces virtually no flux below the Lyman limit or at the 
Lyman transitions of H\,I (see Fig.\,6). 
Any hydrogen gas would thus only follow the Keplerian rotation and be radially stationary
and, as such, could act as the braking agent (see Table\,\ref{rad_press}).

We apply a simplified analysis, including gravity, radiation pressure and gas `friction', 
to examine this possibility. The interaction of the gas with the 
dust particles in the disk is neglected (only big grains with $\beta \ll 1$ and strong coupling
to the gas could decelarate the atoms). The equation of motion of the atom reads

\begin{equation}
m\,\frac{{\rm d}v}{{\rm d}t} = - F_{\rm grav} + F_{\rm rad} - 
                                      F_{\rm fric}\,\,\,.
\end{equation}

Letting $F_{\rm rad} = \beta\,F_{\rm grav}$ and assuming $F_{\rm fric}=C\,v$, 
this reduces to

\begin{equation}
m\,\frac{{\rm d}v}{{\rm d}t} = (\beta - 1)\,F_{\rm grav} - C\,v\,\,\,.
\end{equation}

The radial solution to Eq.\,4 is

\begin{equation}
v = v_{\infty} + (v_0 - v_{\infty})\,e^{ - C\,t/m }
\end{equation}

with $v_0 = v(t=t_0)$ and, as $t \rightarrow \infty$,

\begin{equation}
v_{\infty} = \frac{\beta - 1}{C}\,F_{\rm grav}\,\,\,.
\end{equation}

For collisions of neutrals with neutral hydrogen atoms, we approximate the friction 
coefficient (full momentum deposition) by

\begin{equation}
C_{\rm nn}  =  \pi\, a_0^2\,m_{\rm H}\,n_{\rm H}\,v_{\rm H}     
            =  a_0^2\,\,(8\,\pi\,k\,m_{\rm H})^{\frac{1}{2}}\,T^{\frac{1}{2}}\,n_{\rm H}, 
\end{equation}

where $a_0$ is the Bohr radius. For ions colliding with neutral hydrogen atoms,
Beust et al. (1989) proposed (hyperbolic orbit approximation) 

\begin{equation}
C_{\rm in} =  \pi\, b_0^2\,m_{\rm H}\,n_{\rm H}\,v_{\rm H} =
              \left ( \pi\,\alpha_{\rm H}\,m_{\rm H}/\epsilon_0  \right )^{\frac{1}{2}} 
              q\,n_{\rm H},
\end{equation}

where $b_0$ is the impact parameter, $\alpha_{\rm H}$ is the polarisability of hydrogen and 
$q$ is the electric charge of the ion.
The characteristic time scale to approach $v_{\infty}$ is given by the $e$-folding time

\begin{equation}
\tau = m/C\,\,\,. 
\end{equation}

For Eq.\,6 to be valid $\tau$ needs to be shorter than the time scale for significant 
changes in the disk 
(e.g., $\mid\!\!{\rm d} F_{\rm grav}/F_{\rm grav}\!\!\mid \sim 2\,v_{\infty}\,\tau/r \ll 1$), 
a requirement which was verified a posteriori to be generally fulfilled. 

In Fig.\,7, the asymptotic velocity $v_{\infty}$ is shown as a function of the 
radial distance from the star, $r$, for three atomic species having different atomic masses and 
very different values of $\beta$ (Table\,\ref{rad_press}). The disk structure is approximated by
power laws, e.g. $n \propto r^p$, where the density law was varied, but the same
temperature distribution was used throughout [power law exponent $-0.3$ and normalisation at 26\,AU, 
i.e. $T(26\,{\rm AU})=110$\,K].
Density distributions are shown for three example values of $p$, i.e. $p=0$ (constant density), 
$p=-1.5$ and $p=-2.5$ (`standard' density distribution). 

For the neutral species $^{23}$Na\,I, an average value of $\beta=250$ was used and in Fig.\,7
the normalisation values at 26\,AU of the hydrogen density [$n(26\,{\rm AU})$ in \cmthree] are
\powten{6} for $p=0$, \powten{7} for $-1.5$ and \powten{8} for $-2.5$.
As an illuminating example, for the ion $^{40}$Ca\,II, the minimum value $\beta_{\rm min}=3$
was selected and the corresponding $n(26\,{\rm AU})$ are \powten{4.5} for $p=0$ and
\powten{5} for both $-1.5$ and $-2.5$. As the radiation pressure coefficient of $^{52}$Cr\,II
is merely a tiny $4 \times 10^{-7}$, only the constant density case [$n(26\,{\rm AU}) = 10^5$\,\cmthree]
is shown as an illustration.

Whatever the actual gas density distribution in the \bpic\ disk might be, it is clear from these
numerical experiments that average densities would need to be in excess of \powten{6}\,\cmthree\ 
in order to meet the limits set by the observations: the observed
spatial distribution of the line profiles are reproduced to within 0.1\,\kms\ by 
a Keplerian model \cite{brandeker2003}. If this gas were roughly spherically 
distributed, as hinted at by the observations perpendicular to the disk,
the total mass would be in the neighbourhood of 
$M_{\rm H} \sim 0.02\,r_{100}^3\,\mu_2\,n_6$\,\mjup\
($r_{100}$ is in units of 100\,AU; $\mu_2$ is the molecular weight in units of one 
particle per two hydrogen nuclei;
$n_6$ is $n_{\rm H}$ in units of \powten{6}\,\cmthree). The implied column density of 
hydrogen $N_{\rm H} \sim 2.5\times 10^{21}\,n_6\,r_{100}$\,\cmtwo\ would suggest this 
gas to be molecular, as the upper limit on atomic hydrogen is a few times \powten{19}\,\cmtwo\
\cite{freudling1995}. It follows that such an \molh\ component would be consistent with the 
observations of Thi et al. (2001) for $r_{100} \sim 2$. Confirmation of these results by re-newed
observations, with e.g. SIRTF, are thus eagerly awaited. 

Significant amounts of \molh\ around \bpic\ are hard to reconcile with
the absence of detectable \molh\ absorption along the line of sight toward the star, 
at the same time as CO absorption has clearly been detected. If real,
the existence of such relatively large amounts of gas in the \bpic\ disk would have
far reaching implications for the gas-dust dynamics and the evolution of the disk 
\cite{takeuchi2001}. However, as already pointed out, this gas component would be insufficient
to build an analogue to the solar system and may simply represent `left-overs'
from a recent planet formation epoch. 

\subsection{Search for Atomic gas in Other Systems}

To this end, we have also searched the disks around HR\,4796A and \epseri\ for atomic gas.
These searches have been unsuccessful. The reasons for this failure are  possibly quite 
different, though. If the amount of gas would scale with the amount of dust, the dustier
disk around HR\,4796A should exhibit an even clearer gas signature than the \bpic\ disk.
However, HR\,4796A is nearly four times more distant and the scattering disk extends to 
less than one arcsec from the central star \cite{schneider1999}, presenting a 
most unfavourable contrast case for ground based observations. 

\epseri\ on the other hand is more than six times closer to the Earth than \bpic\ and its circumstellar
disk subtends several tens of arcsec in the sky (at sub-millimeter wavelengths). 
It should thus, in principle, be easier to observe. The disk contains merely one percent of 
dust compared to the \bpic\ disk (see also Fig.\,1). Again, 
scaling the gas emission with the dust emission would imply two orders of magnitude 
lower intensities, which would have escaped detection at our present sensitivity.

\section{Conclusions}

Below, the main conclusions of this work are  briefly summarised.

\begin{itemize}
\item[$\bullet$] Searches for molecular gas in dusty disks around main-sequence stars 
have so far been unsuccessful, with the possible exception of the \bpic\ system, 
toward which Thi et al. (2001) reported the discovery of substantial amounts of \molh\ gas.
\item[$\bullet$] Searches for atomic gas in a few of these systems have resulted 
thus far in clear detections of line emission from the  \bpic\ disk \cite{olofsson2001}. 
Follow-up observations revealed the presence of a large number of spectral lines
from neutral and singly ionised non-volatile species \cite{brandeker2003}.
\item[$\bullet$] The results from spatio-spectral fitting of these lines are consistent 
with gas orbiting the star at Keplerian velocities. Deviations from Keplerian motion are 
less than 0.5\,\kms\ at the $5\sigma$ level.
\item[$\bullet$] The analysis of these motions invoking gravity, radiation pressure and 
gas friction on the atoms/ions indicates that some kind of decelerating material must be present
in the disk.
\item[$\bullet$] Identifying this braking agent with hydrogen gas leads to the conclusion that 
average hydrogen densities are $n_{\rm H} > 10^6$\,\cmthree\ and that the hydrogen is most 
likely molecular.
\item[$\bullet$] Without stretching parameters too far, this result could be in support of the 
result by Thi et al. (2001), i.e. that the present amount of gas around \bpic\ is of the order 
of 50\,\mearth.
\end{itemize}

\section*{Acknowledgements}

I wish to thank the organisers for inviting me to this very stimulating conference and
Alexis Brandeker for Figures 2 through 5 and for valuable comments on the manuscript. 

I'd like also to mention that I am convinced that
the collaboration between ESA and NASA will result in a highly successful Darwin/TPF mission.

\newpage

\begin{center}
{\bf Figure Captions}
\end{center}

Fig.\,1. The ratio of integrated excess to photospheric flux as a function of 
  estimates of the stellar age. The most prominent Vega-excess stars are identified. 
  Open circles refer to stellar members of a number of
  open clusters. The point for the Sun is a lower limit, as it traces only the fraction
  of the zodiacal dust inside the Kuiper Belt. Adapted from Spangler et al. (2001).

Fig.\,2. Observations of the \bpic\ disk in the NaI\,D$_2\,\lambda 5890$ line (ESO, Chile). 
  In the right panel of the
  {\bf Left Figure}, the EMMI ($R\sim 60\,000$) observations with the 3.5\,m NTT are shown
  (horizontal scale in \kms, vertical scale in AU). A model of the 
  line emission from a disk in Keplerian rotation is shown to the left, whereas, 
  in the middle, the model is shown degraded to the resolution of the observations, with 
  white noise added and the central stellar spectrum removed \cite{olofsson2001}.
  Follow-up observations with UVES ($R\sim 90\,000$) at the 8\,m VLT, shown in the
  {\bf Right Figure} to scale, largely improved on both the spectral and spatial resolution 
  and coverage \cite{brandeker2003}. Telluric D$_2$ emission is seen at
  $-30$\,\kms\ and the approaching southwest side is up (cf. Fig.\,3). 

Fig.\,3. The slit positions of our VLT-UVES observations are shown superposed onto the
  HST-STIS image of the scattering dust disk \cite{heap2000} projected onto the sky. The
  width of the slit is \asecdot{0}{3} and, during the observations, the seeing was in 
  the range \asecdot{0}{4} to \asecdot{0}{6}. The length of the slit, limited by the
  Echelle inter-order separation, is 8\asec. The observations along the disk were performed
  with overlapping slit positions.

Fig.\,4. Scale heights of the gas disk in three emission lines (ground state transitions).
  The bars show the Full Width to Half Maximum of the line emission at four positions
  in the disk. $x$- and $y$-offsets, in arcsec, are relative to the star \bpic\
  and to the midplane of the scattering dust disk \cite{kalas1995}, respectively. For
  comparison, the dust disk \cite{heap2000} is shown to scale in the lower half of the figure.

Fig.\,5. The spatial distribution of the NaI\,D$_2$ emission in the \bpic\ disk, with the
  NE and SW disk halves identified \cite{brandeker2003}. Error bars represent $1\sigma$ of
  the quadratically summed photon noise of the line and off-line fluxes.
  Radial distances from the star are expressed in arcsec and integrated surface intensities 
  in \ecsarc. The vertical dashed lines refer to the positions of the perpendicular slits.

Fig.\,6. The UV spectrum of $\beta$\,Pic. Continuum points are from FUSE and HST 
  observations (archive data). Also shown are the results of HST-STIS observations
  of the NE- and SW-disk of \bpic\ at \asecdot{0}{6} distance from the star (archive data).
  The line spectrum is based on a stellar model atmosphere (at 19.3\,pc) with $T_{\rm eff}=8000$\,K, 
  $\log g=4.5$ and $\log Z/Z_{\odot}=0.0$ \cite{hauschildt1999}; for the assumed  
  $R=1.75$\,\rsun, the luminosity of the star is $L=11$\,\lsun. 
  The ionisation edges of species discussed in the text are indicated by the vertical bars.

Fig.\,7. The asymptotic velocity $v_{\infty}$ versus the radial distance 
  $r$ from the star. For \textcolor{red}{Na\,I} (A = 23) an average $\beta = 250$ has been used, 
  and the \textcolor{red}{red curves} are for $n(26\,{\rm AU})=10^6,\,10^7\,\,{\rm and}\,\,10^8$\,\cmthree\ for
  $p=0, -1.5\,\,{\rm and}\,\,-2.5$ (dotted, dashed and solid lines) respectively (see the text). 
  For \textcolor{blue}{Ca\,II} (A = 40) $\beta_{\rm min} = 3$ has been applied as
  a strict lower limit to the radiation pressure coefficient (see Table\,\ref{rad_press}). 
  The \textcolor{blue}{blue curves} are for $n(26\,{\rm AU})=10^{4.5}$\,\cmthree\ for $p=0$ and 
  $10^5$\,\cmthree\ for both $p= -1.5$ and $-2.5$.
  For \textcolor{green}{Cr\,II} (A = 52), only the constant density case (\powten{5}\,\cmthree)
  is shown by the \textcolor{green}{green curve}.  
  As this ion is insensitive to radiation pressure, relative velocities are negative. 
  The \textcolor{magenta}{$5\sigma$ upper limit} to the observed excess velocities over Keplerian motion 
  is shown by the \textcolor{magenta}{purple} horizontal line.

\end{document}